\documentclass[12pt]{article}
\usepackage[english]{babel}
\usepackage{amssymb,latexsym,epsfig,cite}
\usepackage{graphicx}
\usepackage{fix-cm}

\newcommand{\be}{\begin{equation}}
\newcommand{\ee}{\end{equation}}
\def\bea{\begin{eqnarray}}
\def\eea{\end{eqnarray}}
\begin{document}

%%%%%%%%%%%%%%%%%%%%%%%%%%%%%%%%%%%%%%%%%%%%%%%%%%

\begin{center}
\Large{\bf $SU(1,1)$ Coherent States For Position-Dependent Mass
Singular Oscillators}
\end{center}

\vskip2ex
\begin{center}
Sara Cruz~y~Cruz${}^1$ and Oscar Rosas-Ortiz${}^2$\\[2ex]
${}^1$ {\em Secci\'on de Estudios de Posgrado e Investigaci\'on, UPIITA-IPN,\\
Av. IPN 2580, CP 07340 M\'exico~DF, Mexico \\[1ex]
${}^2$ Departamento de F\'{\i}sica, Cinvestav,\\
AP 14-740, 07000 M\'exico~DF, Mexico}
\end{center}

\begin{center}
\begin{minipage}{14cm}
{\bf Abstract} The Schr\"odinger equation for position-dependent
mass singular oscillators is solved by means of the factorization
method and point transformations. These systems share their
spectrum with the conventional singular oscillator. Ladder
operators are constructed to close the $su(1,1)$ Lie algebra and
the involved point transformations are shown to preserve the
structure of the Barut-Girardello and Perelomov coherent states.
\end{minipage}
\end{center}

%%%%%%%%%%%%%%%%%%%%%%%%%%%%%%%%%%%%%%%%%%%%%%%%%%%%%%%%%%%%%%%
\section{Introduction}
\label{intro}

There has been much interest in the construction of coherent states
since the origin of quantum mechanics \cite{Sch26}. Following
Glauber, the concept is useful in the quantum description of the
correlation and coherence properties of light \cite{Gla06}. The
Glauber states deal with the dynamical properties of the harmonic
oscillator and posses three basic properties: they (1) are
eigenfunctions of the annihilation operator belonging to complex
eigenvalues, (2) are displaced versions of the ground state
wave-function and (3) minimize the uncertainty relation between
position and momentum. For systems other than the harmonic
oscillator, such properties are used as different definitions of the
coherent states and, in general, they are not equivalent to each
other. Thus, the term {\em coherent states} (CS) has been used for a
wide class of mathematical objects over the years. For instance, the
generalized CS studied by Barut and Girardello \cite{Bar71} and by
Perelomov \cite{Per72} are based on the first and second properties
of the Glauber states. In turn, the construction of CS as minimizing
the uncertainty relation of a pair of observables is incidentally
found as a secondary result for some special systems. Nevertheless,
the most valuable property of the CS is that they can be studied for
many systems in terms of the definition leading to the desirable
result.

The purpose of this work is to draw the CS applications to the
study of position-dependent mass (PDM) systems in quantum
mechanics. Some previous results include the CS for the harmonic
\cite{Cru09,Ju,Biswas,Chi} and nonlinear \cite{Midya,Chi}
oscillators. We are interested in the construction of PDM
Hamiltonians such that their spectrum is exactly the same as that
of the conventional singular oscillator. We shall show that the
description of these new systems obeys the $su(1,1)$ algebra. The
structure of the paper is as follows. In Section~\ref{gral},
expressions for the mappings and the operators leading to the
relevant algebras are given. The algebraic structure of the PDM
linear oscillators is revisited in Section~\ref{position}.
Departing from the ladder operators of the linear case, in
Section~\ref{singpot} we construct a new class of operators and
show that they are connected in a natural way to the singular
oscillator. It is shown that these operators are the generators of
the $su(1,1)$ Lie algebra (the subject studied also in
\cite{Roy05}). In Section \ref{appl} some applications are given
and the corresponding CS are constructed \emph{a la}
Barut-Girardello and \emph{a la} Perelomov. Some final remarks
close the paper.

%%%%%%%%%%%%%%%%%%%%%%%%%%%%%%%%%%%%%%%%%%%%%%%%%%%%%%%%%%%%%
\section{Factorization and point transformations}
\label{gral}

Consider a one-dimensional quantum system of mass $m(x)$ acted by
the potential $V(x)$. A convenient expression for the Hamiltonian
reads
\be
H_a = \frac12 m^a P m^{2b} P m^a + V, \qquad 2a+2b =-1
\label{hamil1}
\ee
where $P$ and the position operator $X$ satisfy
$[X,P]=i\hbar$. This Hamiltonian admits the factorization
\be
H_a = AB + \epsilon, \qquad A = B^{\dagger} = -\frac{i}{\sqrt 2}
m^a P m^b + \beta,
\label{factor1}
\ee
with $\epsilon$ a constant (in energy units) to be fixed. In the
position representation ($X=x$ and $P=-i\hbar \frac{d}{dx}$) the
function $\beta$ is a root of the Riccati equation:
\be
V -\epsilon = \frac{\hbar}{\sqrt{2m}} \left[ 2\left( a +
\frac{1}{4} \right) \left( \frac{m'}{m} \right) \beta -
\beta' \right] + \beta^2, \qquad ' \equiv \frac{d}{dx},
\label{riccati1}
\ee
and the factorization operators satisfy the commutation rule
\be
[A,B] = -\frac{\hbar^2}{m^3} \left(a+\frac{1}{4}\right) \left[
mm'' - \frac{3(m')^2}{2} \right] - \frac{2\hbar}{\sqrt{2m}}
\beta'.
\label{conmuta1}
\ee
The PDM problem can be simplified using the map $\psi \rightarrow
e^g \varphi$ and the change of the independent variable $x$, ruled
as follows:
\be
x \mapsto y=s(x), \qquad y \mapsto x=s^{-1}(y).
\label{indep1}
\ee
To define $s$ as a bijection we shall assume that $J\left({\cal
D}_a\right)\neq 0$, with ${\cal D}_a \equiv Dom (H_a)$ and $J= s'$
(see \emph{e.g.} \cite{Cho77}). The combination of transformations
(\ref{indep1}) produces
\be
y = \int J(x) dx + y_0, \qquad J(x) =e^{2g(x)} = \left[
\frac{m(x)}{m_0} \right]^{1/2}
\label{mapping1}
\ee
and
\be
\psi(x) = J^{1/2}(x) \varphi_*(y)
\label{mapping2}
\ee
where $m_0$ and $y_0$ are integration constants in proper units
(hereafter we take $y_0=0$). The function $\varphi_*$ is the
representation of $\varphi$ in the $y$-space: $\varphi(x) =
\varphi(s^{-1}(y)) = [\varphi \circ s^{-1}](y) \equiv
\varphi_*(y)$ and viceversa $\varphi_*\circ s =\varphi$. A further
simplification (see \cite{Cru09} for details) leads to the
Schr\"{o}dinger equation
\be
H_*^{(a)} \varphi_* (y) :=\left[-\left(
\frac{\hbar^2}{2m_0}\right) \frac{d^2}{dy^2} + V_*(y)\right]
\varphi_*(y) =E \varphi_*(y),
\label{eigen4}
\ee
with ${\cal D}^{(a)}_* = \textrm{Dom} (H_*^{(a)})$ and
$\textrm{Sp} (H_a) =
\textrm{Sp}(H_*^{(a)})$. Hence two general cases are
distinguishable:

%%%%%%%%%%%%%%%%%%%%%%%%%%%%%%%
\vskip1ex
$\bullet$ {\bf MDNT} (Mass-dependent null terms) The
mass-function $m(x)$ is a solution of the non-linear, second order
differential equation
\be
m m'' -\left(\frac 74 + a\right) (m')^2=0.
\label{mass1}
\ee
A simple analysis shows that the roots of (\ref{mass1}) leading to
appropriate bijections $s$ have the form
\begin{equation}
\label{massbi}
m(x;a) = m_0 \left(x_0+ \lambda x\right)^{-4/(3+4a)}
\end{equation}
with $a$ in the set
\be
{\cal A} = \left\{ a_0 =-1/4, a_n =
\frac{1-n}{4n} \right\}, \quad n \in \mathbb{N},
\label{set3}
\ee
and $x_0$ a dimensionless real constant while $m_0$ and $\lambda$
are real constants expressed in mass and inverse of distance units
respectively. Then the functions
\be
m_{(0)}(x) \equiv m(x;a_0) =
\frac{m_0}{(x_0 + \lambda x)^2}, \quad s_{(0)} (x)= \frac{\ln (x_0
+ \lambda x)}{\lambda}, \quad x \geq t_0 =-\frac{x_0}{\lambda},
\label{m14}
\ee
define an invertible mapping from ${\cal D}_{a_0} \subseteq
[t_0,+\infty)$ to ${\cal D}_*^{(a_0)} \subseteq \mathbb R$ and the
pair
\be
m_{(n)}(x) \equiv m(x;a_n)= m_0 (x_0+\lambda
x)^{-\frac{4n}{(2n+1)}}, \qquad s_{(n)} (x) = \frac{(x_0 + \lambda
x)^{\frac{1}{(2n+1)}}}{\lambda (2n+1)^{-1}}, \quad n\in \mathbb{N}
\label{msing2}
\ee
corresponds to a bijection between ${\cal D}_{a_n} \subseteq
\mathbb R$ and ${\cal D}_*^{(a_n)} \subseteq \mathbb R$.

%%%%%%%%%%%%%%%%%%%%%%%%%%%%%%%%%%%%%%%

\vskip1ex
$\bullet$ {\bf MINT} (Mass-independent null terms) Given
a properly defined mass-function $m(x)$, the ordering of $P$ and
$m(x)$ in (\ref{hamil1}) is {\it a priori\/} fixed as $a=b=-1/4$.
For instance, the regular functions
\be
m_R(x) = \frac{m_0}{1+(\lambda x)^2}, \qquad s_R(x) = \frac{{\rm
arcsinh}(\lambda x)}{\lambda}, \qquad \lambda \in \mathbb{R}
\label{mass}
\ee
define a bijection connecting ${\cal D}_{-1/4} \subseteq \mathbb
R$ to ${\cal D}^{(-1/4)}_* \subseteq \mathbb R$. Another well
behaved bijection (see Section \ref{appl}) is given by
\be
m_e(x)=m_0 e^{2\lambda x}, \qquad s_e(x)= \frac{e^{\lambda x}
-x_0}{\lambda},
\label{newmass}
\ee
and connects ${\cal D}_{-1/4} \subseteq \mathbb R$ to ${\cal
D}^{(-1/4)}_* \subseteq [t_0,+\infty)$.

%%%%%%%%%%%%%%%%%%%%%%%%%%%%%%%%%%%%%%%%%%%%
\section{Position-dependent mass linear oscillators}
\label{position}

Let the commutator (\ref{conmuta1}) be a constant, namely
$[A,B]=-\hbar \omega_0$, with $\omega_0$ in frequency units. The
$\beta$-function is easily found to be
\be
\beta = \frac{\omega_0}{\sqrt 2} \int^x m^{1/2} dr - \frac{\hbar}{\sqrt 2}
\left( a + \frac{1}{4} \right) \left(\frac{m'}{m^{3/2}}\right) +
\beta_0
\label{beta}
\ee
with $\beta_0$ an integration constant which will be omitted in
the sequel. The identification $\epsilon = \hbar \omega_0/2$,
after introducing (\ref{beta}) in the Riccati equation
(\ref{riccati1}), leads to a very simple form of the potential in
the $y$-representation
\be
V(x)= \frac{\omega_0^2}{2} \left[ \int^x m^{1/2} dr \right]^2
=\frac{m_0\omega_0^2}{2} \left[ \int^x J dr \right]^2 = \left(
\frac{m_0\omega_0^2}{2}\right) y^2 \equiv V_*(y).
\label{pot2}
\ee
Thereby the transformation (\ref{mapping1})-(\ref{mapping2}) tunes
to the potentials $V(x)$ exhibiting the equidistant energies
$\hbar \omega_0 (n+1/2)$ of a constant mass quantum oscillator.
Since $s$ is a bijection, (\ref{pot2}) admits another lecture:
given $V_*(y)=\frac{m_0\omega_0^2}{2} y^2$, the mapping
(\ref{mapping1})-(\ref{mapping2}) leads to a potential
$V(x)=\frac{m_0\omega_0^2}{2} (s(x))^2$ which is isospectral to
the harmonic oscillator for the masses $m(x)$ allowed by the rule
$s$.

It is convenient to introduce a dimensionless notation by taking
$[E]=\hbar \omega_0$ and $[L]=\sqrt{\hbar/(m_0 \omega_0)}$ as the
units of energy and distance respectively. Hence, the potential
(\ref{pot2}) reads $V_*(y) = \frac12 \textrm{y}^2 [E]$, with
$\textrm{y}$ a real number such that $y=\textrm{y} [L]$. From now
on we drop the dimensions and use the same symbol for the physical
and the dimensionless variables. We shall return to the
expressions with units only if necessary. In each case, the
notation will be self-consistent. The same holds for the subscript
``*'' labelling the representation of functions and operators in
the $y$-space.

The introduction of (\ref{beta}) in (\ref{factor1}) cancels the
explicit dependence of the factorization operators $A$ and $B$ on
the ordering label $a$. We have
\be
A= a_+ + \frac12 \left(\frac{d}{dy} \ln J\right), \qquad B= a_- -
\frac12 \left(\frac{d}{dy} \ln J\right), \qquad
[A]=[B]=\sqrt{[E]/2}
\label{factors}
\ee
with $a_-$ ($a_+$) the conventional annihilation (creation)
operator of the quantum oscillator in the $y$-space
\be
a_+^{\dagger} = a_- := \frac{d}{dy} + y, \quad [a_-,a_+]=2, \quad
a_+ a_- = 2N, \quad [2N,a_{\pm}]=\pm 2a_{\pm}.
\label{aniq}
\ee
Here $N$ is the Fock's number operator. A most convenient
relationship between $A$, $B$ and $a_{\pm}$ is easily calculated
to read
\be
AJ^{1/2} =J^{1/2} a_+, \qquad BJ^{1/2} =J^{1/2} a_-.
\label{abes}
\ee
Hence, the action of $A$ and $B$ in the $\varphi$-space can be
established as
\be
A \psi = \left(J^{1/2}a_+\right) \varphi,
\qquad B \psi = \left(J^{1/2} a_-\right) \varphi.
\label{actions1}
\ee
The Hamiltonian $H_a$ is clearly isospectral to the
one-dimensional quantum oscillator
\be
\begin{array}{rl}
H_a \psi & = (AB+1)\psi = A \left(J^{1/2} a_-\right)\varphi + \psi=
J^{1/2}(2N+1)\varphi\\[1ex]
& = J^{1/2} \left[-\frac{d^2}{dy^2} + y^2 \right]\varphi = J^{1/2}
H^{(a)} \varphi= (2n+1) \psi,
\end{array}
\label{actions2}
\ee
where $[H_a]= [H^{(a)}]= [E]/2$.

%%%%%%%%%%%%%%%%%%%%%%%%%%%%%%%%%%%%%%%%%%%%%%%%%%%%%%%%%
\section{Position-dependent mass singular oscillators}
\label{singpot}

Some consequences of the commutation relations (\ref{aniq}) are that
\be
[a_-,f]= f_y, \quad [a_+,f]= -f_y, \quad [2N,f] = -f_{yy} - 2f_y
\frac{d}{dy}, \quad  f_y \equiv \frac{df}{dy}
\label{comm1}
\ee
with $f$ a differentiable function of the position. Moreover,
since $[a_-^2, a_+^2]=8H^{(a)}$ we can introduce the operators
\be
c_+:= a_+^2+f, \qquad c_-:= a_-^2+f
\label{ces}
\ee
to get
\be
[c_-,c_+]=8 \left[ H^{(a)}+\left( \frac{y}{2}\right) f_y \right]:=
8h^{(a)}.
\label{comm2}
\ee
The straightforward calculation shows that given
$f(y)=-\frac{g_0}{2y^2}$, with $g_0$ a real constant, the
operators $k_0=h^{(a)}/4$, $k_\pm = c_\pm/4$ close the $su(1,1)$
algebra
\be
[k_-,k_+]=2k_0, \qquad [k_0,k_{\pm}]=\pm k_{\pm}.
\label{su11}
\ee
The operator $h^{(a)}$ in (\ref{comm2}) is the Hamiltonian of the
singular oscillator:
\be
h^{(a)} = H^{(a)} + \frac{g_0}{2y^2} = -\frac{d^2}{dy^2} +
y^2 + \frac{g_0}{2y^2}, \qquad [g_0]=[L]^4.
\label{hsing}
\ee
Remark that the potential $V(y)= y^2 + \frac{g_0}{2y^2}$ admits an
infinite point spectrum if $g_0>-1/2$ (see e.g. Chs. III.18 and
V.35 of \cite{Lan60}). Moreover, the presence of the
centrifugal-like term $\frac{g_0}{2y^2}$ constrains the domain of
definition of $h^{(a)}$ to be ${\cal D}^{(a)}_* =[0,+\infty)$, and
the wave-functions of $h^{(a)}$ are necessarily equal to zero at
the origin. In this way, if $g_0=0$, only the odd
linear-oscillator functions are recovered (see
Section~\ref{phys}). On the other hand, from (\ref{abes}) and
(\ref{actions2}) one obtains the expressions for the generators in
the PDM case:
\be
C_+ = A^2 -\frac{g_0}{2s^2(x)}, \qquad  C_-= B^2
-\frac{g_0}{2s^2(x)}, \qquad h_a = H_a + \frac{g_0}{2s^2(x)}.
\label{su11new}
\ee
Notice that the relationships
\be
C_{\pm} J^{1/2} =J^{1/2}c_{\pm}, \qquad h_a J^{1/2} = J^{1/2}
h^{(a)},
\label{relatito}
\ee
imply that the operators (\ref{su11new}) also close the $su(1,1)$
algebra provided that $K_0=h_a/4$ and $K_\pm = C_\pm/4$.

%%%%%%%%%%%%%%%%%%%%%%%%%%%%%%%%%%%%%%%%%%%%%%%
\subsection{Physical Solutions}
\label{phys}

Let us introduce the mappings $\varphi \rightarrow y^{\ell}
e^{-y^2/2} u(y)$ and $y^2 \rightarrow z$ so that the eigenvalue
problem $h^{(a)} \varphi = E\varphi$ is mapped to the Kummer
equation \cite{Neg00}:
\be
zu_{zz} + \left( \ell + \frac12 -z\right) u_z -\frac{1}{4}
\left(2\ell +1 -E\right)u =0,\qquad \ell^2 -\ell -
\frac{g_0}{2}=0.
\label{kummer}
\ee
Then, for each $\ell_{\pm} = \alpha_{\pm}=\frac12\left(1\pm
\sqrt{1+2g_0}\right)$, we have a general expression of the form
\be
\varphi_{\pm} = \lambda^{(1)}_{\pm} y^{\alpha_{\pm}} e^{-y^2/2}
{}_1F_1 (a,c,y^2) +\lambda^{(2)}_{\pm} y^{1-\alpha_{\pm}}
e^{-y^2/2} {}_1F_1 (\widetilde{a},\widetilde{c},y^2),
\label{general}
\ee
with $4a= 2\alpha_{\pm} +1 -E$, $2c=2\alpha_{\pm} +1$,
$4\widetilde{a}= 3-2\alpha_{\pm} -E$ and $2\widetilde{c}= 3-
2\alpha_{\pm}$. The straightforward calculation shows that both of
the general expressions (\ref{general}) lead to the same physical
solution if $-\frac12 < g_0$. In such a case we omit $\varphi_-$
and take $\lambda^{(2)}_+=0$,  $a=-n$ and $\alpha_+ =\alpha$ to
get
\be
\begin{array}{rl}
\varphi_n(y) & = c_n y^{\alpha} e^{-y^2/2} {}_1F_1
\left(-n,\alpha+\frac12, y^2 \right)\\[2ex]
& = \left( \frac{2n!}{\Gamma(\alpha+n+1/2)}\right)^{1/2} y^{\alpha}
e^{-y^2/2} L_n^{(\alpha-1/2)} (y^2), \quad n=0,1,2,\ldots
\end{array}
\label{physical}
\ee
where $L_n^{(\gamma)} (x)$ are the {\em Generalized Laguerre
Polynomials} \cite{Abr72}. The set of energies is then defined by
\be
E_n=4n+2+\sqrt{1+2g_0} = 4n+2\alpha +1=4(\kappa+n) , \qquad n
=0,1,2,\ldots
\label{physical2}
\ee
The change $y \mapsto y/\sqrt{2}$ in $h^{(a)}$ makes clear that
the set (\ref{physical2}) corresponds to the quantum oscillator
spectrum $E_n=2n+1$, shifted by $\alpha-1/2$. This last case has
been discussed in \cite{Cru09} for $g_0=1/2$. On the other hand,
if $g_0=0$ then $\alpha=1$ and we have
\be
\varphi_n(y)  = \frac{1}{\pi^4 \sqrt{(2n+1)!}} e^{-y^2/2}
H_{\textrm{e}2n+1} (\sqrt{2} y), \quad n=0,1,2,\ldots
\label{physical3}
\ee
with $H_{\textrm{e}2n+1}(x)$ the odd {\em Hermite Polynomials}
\cite{Abr72} and $E_n=4n+3$. That is, the wave-functions
(\ref{physical}) are reduced to the odd-oscillator ones as $g_0
\rightarrow 0$.

%%%%%%%%%%%%%%%%%%%%%%%%%%%%%%%%%%%%%%%%%%%%%%%

\section{Applications}
\label{appl}

Each of the pairs $(V(x),m(x))$, with $V(x)=s^2(x) + g_0/2s^2(x)$
describe a PDM quantum system sharing its spectrum with a particle
of mass $m_0$ subject to the singular oscillator interaction. We
give some examples below.

%%%%%%%%%%%%%%%%%%%%%%%%%%%%%
\vskip1ex
$\bullet$ {\bf MDNT} For $a = a_0 \in {\cal A}$, we have
\be
V_1(x)= \frac{\ln^2(x_0+\lambda x)}{\lambda^2} + \frac{g_0
\lambda^2}{2 \ln^2(x_0+\lambda x))},
\label{potx1}
\ee
with (\ref{m14}) the mapping from ${\cal D}_{a_0}$ to ${\cal
D}^{(a_0)}_* =[0,+\infty)$ and viceversa. On the other hand, if
$a=a_n \in {\cal A}$, $n\in \mathbb{N}$, the potential is given by
\be
V_2(x)= \left(\frac{2n+1}{\lambda}\right)^2 (x_0+\lambda
x)^{2/(2n+1)} + \frac{g_0}{2}\left(\frac{\lambda}{2n+1}\right)^2
(x_0+\lambda x)^{-2/(2n+1)}
\label{potx2}
\ee
with ${\cal D}_{a_n} =[t_0,+\infty)$ and ${\cal D}^{(a_n)}_*
=[0,+\infty)$ connected by (\ref{msing2}).

\begin{figure}[h]
\centering
\includegraphics[width=5.7cm]{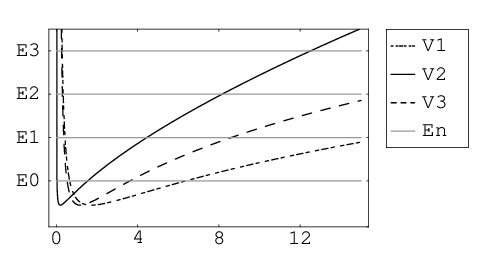} \hskip.5cm
\includegraphics[width=5.7cm]{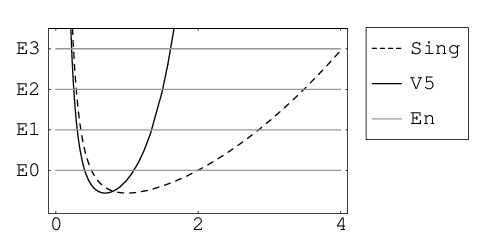}

\caption{\footnotesize The PDM potentials
(\ref{potx1})--(\ref{pot4xa}), together with the singular
oscillator of constant mass (Sing). Potential $V_2(x)$ has been
depicted with $n=1$. The first four energy levels (En) are
included as reference. In all cases $g_0=2$, $\lambda=1$ and $x_0$
is fixed to give ${\cal D}_a=[0,+\infty)$. {\em Vertical} and {\em
horizontal axis} are respectively in $\hbar \omega_0/2$ and
dimensionless units.}
\label{fig1}
\end{figure}

\begin{figure}
\centering
\includegraphics[width=5cm]{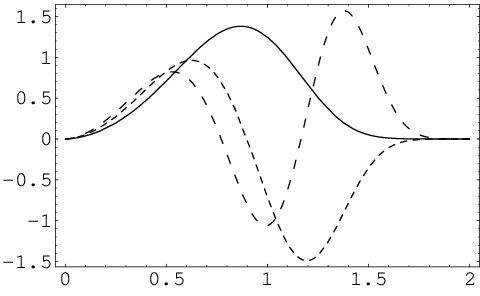} \hskip.5cm
\includegraphics[width=6.5cm]{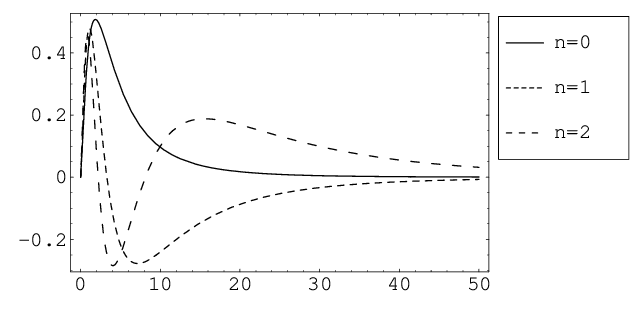}

\caption{\footnotesize The first three wave-functions of the PDM
potentials $V_4(x)$--left and $V_1(x)$ right--for the same values
of the parameters as in Figure~\ref{fig1}. {\em Vertical} and {\em
horizontal axis} are in dimensionless units.} \label{fig2}
\end{figure}

\begin{figure}
\centering
\includegraphics[width=5.2cm]{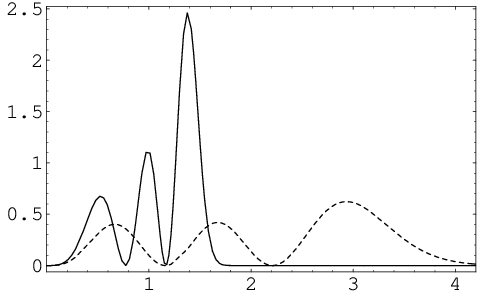}

\caption{\footnotesize The second excited state probability
density of the PDM potential (\ref{pot4xa}) and the singular
oscillator of constant mass ({\em dashed}) for the parameters
indicated in Figure~\ref{fig1}. {\em Vertical} and {\em horizontal
axis} are in dimensionless units.} \label{fig3}
\end{figure}

%%%%%%%%%%%%%%%%%%%%%%%%%%%%%%%%
\vskip1ex
$\bullet$ {\bf MINT} If we fix $a=-1/4$, the transformation
(\ref{mass}) connects ${\cal D}_{-1/4}=[0,+\infty)$ with ${\cal
D}^{(-1/4)}_*=[0,+\infty)$ and leads to the potential
\be
V_3(x) = \frac{\textrm{arcsinh}^2(\lambda x)}{\lambda^2} +
\frac{g_0 \lambda^2}{2 \textrm{arcsinh}^2(\lambda x)}.
\label{potx3}
\ee
In turn, for $x_0 =1$, transformation (\ref{newmass}) defines the
mapping from ${\cal D}_{-1/4} = [0, +\infty)$ to ${\cal
D}^{(-1/4)}_* =[0,+\infty)$ and the potential
\be
V_4(x)= \frac{4 e^{\lambda x}}{\lambda^2} \sinh^2 \left(
\frac{\lambda x}{2} \right) + \left(\frac{g_0\lambda^2}{8}\right)
e^{-\lambda x}  \,
\textrm{csch}^2 \left( \frac{\lambda x}{2} \right), \qquad x>0.
\label{pot4xa}
\ee
In Fig.~\ref{fig1} the global behavior of the above defined
potentials is shown for specific values of the parameters.
Fig.~\ref{fig2} shows the eigenfunctions of potentials $V_1(x)$
and $V_4(x)$. In the former case the functions are expanded
towards infinity by preserving the shape of the constant-mass ones
at short distances. In the second case they are squeezed into the
vicinity of the origin of coordinates. As expected, no change in
the normalization is found. Details can be appreciated in
Fig.~\ref{fig3} where one of the squeezed probability densities of
$V_4(x)$ is contrasted with its constant-mass equivalent. As a
final remark, the above results make clear that $\kappa = \frac12
(\frac12 + \alpha)$ determines the representation of $SU(1,1)$ we
are dealing with (see \emph{e.g.} \cite{Bar77} and \cite{Bri96}).

%%%%%%%%%%%%%%%%%%%%%%%%%%%%%%%%%
%%%%%%%%%%%%%%%%%%%%%%%%%%%%%%%%%%%%%

%%%%%%%%%%%%%%%%%%%%%%%%%%%%%%%%%
%%%%%%%%%%%%%%%%%%%%%%%%%%%%%%%%%%%%

%%%%%%%%%%%%%%%%%%%%%%%%%%%%%%%%%
%%%%%%%%%%%%%%%%%%%%%%%%%%%%%%%%%%%%%

%%%%%%%%%%%%%%%%%%%%%%%%%%%%%%%%%%%%%%%%%%%%%%%%%%%%%%%%%%

\subsection{New $SU(1,1)$ coherent states}
\label{newcs}

Similar to the case of $c_+$, $c_-$ and $h^{(a)}$, the operators
$C_{\pm}$ intertwine the Hamiltonian $h_a$ with itself, shifted by 4
units of the energy. In other words, $C_{\pm}$ work as ladder
operators when acting on the wave-functions of $h_a$. We get
$C_-\psi_0=0$ and
\be
C_{\pm} \psi_n = \gamma_{\pm}(n) \psi_{n\pm 1}, \qquad
\gamma_{\pm} (n)= \sqrt{E_{n\pm 1} E_n +3-2g_0}.
\label{accmas}
\ee
It is convenient to rewrite the coefficients $\gamma_{\pm}$ as
follows
\be
\gamma_+(n) =4 \sqrt{(n+1)(n+ 2\kappa)}, \qquad
\gamma_-(n) = 4 \sqrt{n(n-1+2\kappa)}.
\label{gamis}
\ee
The $su(1,1)$ CS for the PDM singular oscillator are now
constructed as solutions of the equation $C_-\Phi_z= z \Phi_z$,
$z\in \mathbb{C}$. We get
\be
\Phi_z (x)= \left( \frac{\vert
z\vert}{4} \right)^{\kappa-1/2} \left[ \frac{J(x)}{I_{2\kappa-1}
(\vert z \vert/2)} \right]^{1/2} \sum_{\ell =0}^{\infty}
\frac{(z/4)^{\ell}}{\sqrt{\ell ! \Gamma(\ell +2\kappa)}}\,
\varphi_{\ell}(s(x))
\label{cobar1}
\ee
where $I_{\nu}(z)$ is the $\nu$-order modified Bessel function of
the first kind \cite{Abr72}. The constant mass case is recovered
by taking $m(x)=m_0$, then $J=1$ and $s(x)=x$. A final change
$z\mapsto 4z$ leads from (\ref{cobar1}) to the well known
generalized coherent states of Barut and Girardello \cite{Bar71}.

In a similar form we get the involved Perelomov $SU(1,1)$ coherent
states:
\be
\widetilde{\Phi}_z(x)  = J^{1/2}(x) (1-\vert
4z\vert^2)^{\kappa} \displaystyle\sum_{\ell=0}^{\infty} (4z)^{\ell}
\left[ \frac{\Gamma( \ell +2\kappa)}{\ell !\Gamma(2\kappa)}
\right]^{1/2} \varphi_{\ell}(s(x))
\label{perez}
\ee
with the same recipe to recover the classical results
\cite{Per72}. In the description of Perelomov, each $SU(1,1)$ CS
is connected with a point in the coset space $SU(1,1)/U(1)$.
Thereby, the construction (\ref{perez}) corresponds to the
applying of the unitary operators $\Omega(\xi) \in SU(1,1)/U(1)$
to the lowest state $\psi_0$. Here $4z =(\xi/\vert \xi \vert)
\tanh \vert \xi \vert$ so that $\vert 4z \vert <1$. The details
can be consulted in \cite{Per72} (see also \cite{Bri96}).

Both of the above derived sets of CS preserve the form of the
constant-mass case. Important properties like the resolution of
the identity are found to be similar to the conventional case.
These states also evolve in time without dispersion because their
energy eigenvalues are equally spaced. Moreover, the quadratures
$\mathcal{X}(X,P)= \left(K_++K_-\right)/2$ and $\mathcal{P}(X,P)=
i\left(K_+-K_-\right)/2$ appear to satisfy the commutator
$\left[\mathcal{X},\mathcal{P}\right]=iK_0$. In this sense, the
Barut-Girardello CS can be considered as the quadrature states
minimizing the inequality relation
\begin{equation}
\label{mini}
\Delta\mathcal{X} \Delta\mathcal{P}\geq \frac18\left|\langle
h_a\rangle\right|.
\end{equation}
Some other PDM coherent states minimizing similar inequality
relations are constructed in \cite{Cru09,Biswas,Ju,Midya,Chi}. The
relevant aspect is that the point transformations analyzed in
Section~\ref{gral} simplify the construction of the PDM coherent
states to be practically the same method as in the constant-mass
case.

%%%%%%%%%%%%%%%%%%%%%%%%%%%%%%%%%%%%%%%%%%%%%%%%%%%%

\section{Concluding remarks}
\label{conclusion}

The Darboux transformation of the singular oscillator with a
varying frequency and constant mass $V(y,t)= \omega^2(t)y^2+
\frac{g_0}{2y^2}$ has been studied in \cite{Sam04}. As a result,
it was shown that the CS belonging to $V(y,t)$ are essentially
unchanged, just as the results reported here for the stationary
case and masses varying with the position $m(x)$. A combination of
these approaches would be applicable in the case of a
time-dependent frequency $\omega(t)$ and a position-dependent mass
$m(x)$. It is then expected a similar result: the Barut-Girardello
and Perelomov CS will preserve their global properties after the
transformations. In this context, it is important to remark that
the $\beta$-function defined in (\ref{beta}) obeys the fact that
the operators $A$ and $B$ are taken to fulfill the oscillator
algebra $[A,B]=-\hbar \omega_0$. That is, the function in
(\ref{beta}) corresponds to a particular solution of the Riccati
equation (\ref{riccati1}). As it is well known, general solutions
give rise to different algebras (see e.g. \cite{Mie04} and
\cite{And04}). These algebras have been applied in the
construction of a new kind of CS connected with the linear
oscillator and its Susy-partners \cite{Fer94,Fer07}. Quite
recently, it has been shown that non-linear Susy algebras can be
linearized to exhibit the Heisenberg-Weyl structure. In
particular, the $SU(1,1)$ algebra as connected with the infinite
well was analyzed \cite{Fer07}. Then, the higher order Susy
transformations can be also studied for the position-dependent
mass systems we have presented in this work. The classical models
of the harmonic oscillator and the P\"oschl-Teller potentials have
been also useful in the solving of mass-dependent systems
\cite{Cru07} and in the construction of CS \cite{Cru08}. Of
particular interest, the study of the Wigner function gives rise
to a better understanding of the PDM coherent states
\cite{Biswas,Souza}. Further insights are in progress.

%%%%%%%%%%%%%%%%%%%%%%%%%%%%%%%%%%%%%%%%%%%%%%%%%%%%%%%%%%%
\vskip2ex
\noindent
{\bf Acknownledgementes} The support of CONACyT project
24333-50766-F and IPN grant COFAA is acknowledged.

%%%%%%%%%%%%%%%%%%%%%%%%%%%%%%%%%%%%%%%%%%%%%%%%%%%%%%%%%%%%%%%%


\begin{thebibliography}{99}

\bibitem{Sch26}
Schr\"odinger E., {\em Naturwissenschaften} {\bf 14} 664 (1926)

\bibitem{Gla06}
Glauber R.J., {\em Chem. Phys. Chem.} {\bf 7} 1618 (2006)

\bibitem{Bar71}
Barut A.O. and Girardello L., {\em Commun. Math. Phys.} {\bf 21} 41
(1971)

\bibitem{Per72}
Perelomov A.M., {\em Commun. Math. Phys.} {\bf 26} 222 (1972);
Perelomov A.M., {\em Usp. Fiz. Nauk} {\bf 123} 23 (1977); Perelomov
A., {\em Generalized Coherent States and Their Applications}
(Springer-Verlag, Heidelberg, 1986)

\bibitem{Cru09}
Cruz~y~Cruz S. and Rosas-Ortiz O., {\em J. Phys. A: Math. Theor.}
\textbf{42} 185205 (2009)

\bibitem{Biswas}
Biswas A. and Roy B., \emph{Mod. Phys. Lett. A} \textbf{24} 1343
(2009)

\bibitem{Ju}
Guo-Xing J., Chang-Ying C., and Zhong-Zhou, R., \emph{Commun. Theor.
Phys.} \textbf{51} 797 (2009)

\bibitem{Chi}
Rubi V.C., and Senthilvelan M., \emph{J. Math. Phys.} \textbf{51}
052106 (2010)

\bibitem{Midya}
Midya B., Roy B. and Biswas A., \emph{Phys. Scr.} \textbf{79} 065003
(2009)

\bibitem{Roy05}
Roy B., \emph{Europhys. Lett.} \textbf{72} 1 (2005)

\bibitem{Cho77}
Choquet-Bruhat Y., DeWitt-Morette C. and Dillard-Bleick M., {\em
Analysis, Manifolds and Physics} (North-Holland, Amstersam, 1977)

\bibitem{Lan60}
Landau L.D. and Lifshitz E.M., {\em Quantum Mechanics} (Pergamon,
London, 1960)

\bibitem{Neg00}
Negro J., Nieto L.M. and Rosas-Ortiz O., {\em J. Math. Phys.} {\bf
41} 7964 (2000); Rosas-Ortiz O., Negro J. and Nieto L.M., {\em Rev.
Mex. Fis.} {\bf 49 S1} 88 (2003)

\bibitem{Abr72}
Abramowitz M. and Stegun I. (Eds), {\em Handbook of Mathematical
Functions} (Dover Pub. Inc., New York, 1972)

\bibitem{Bar77}
Barut A.O. and Raczka R., {\em Theory of Group Representations And
Applications} (Polish Scientific Publishers, Poland, 1977)

\bibitem{Bri96}
Brif C. Vourdas A. and Mann A., {\em J. Phys. A: Math. Gen.} {\bf
29} 5873 (1996)

\bibitem{Sam04}
Samsonov B.F., {\em J. Phys. A: Math. Gen.} {\bf 37} 10273 (2004)

\bibitem{Mie04}
Mielnik B. and Rosas-Ortiz O., {\em J Phys A: Math Gen} {\bf 37}
10007 (2004)

\bibitem{And04}
Andrianov A.A. and Cannata F., {\em J Phys A: Math Gen} {\bf 37}
10297 (2004)

\bibitem{Fer94}
Fern\'andez D., Hussin V. and Nieto L.M., {\em J. Phys. A: Math.
Gen.} {\bf 27} 3547 (1994); Fern\'andez D., Nieto L.M. and
Rosas-Ortiz O., {\em J. Phys. A: Math. Gen.} {\bf 28} 2693 (1995);
Rosas-Ortiz O., {\em J. Phys. A: Math. Gen.} {\bf 29} 3281 (1996);
Fern\'andez D. and Hussin V., {\em J. Phys. A: Math. Gen.} {\bf 32}
3603 (1999)

\bibitem{Fer07}
Fern\'andez D., Hussin V. and Rosas-Ortiz O., {\em J. Phys. A: Math.
Gen.} {\bf 40} 6491 (2007); Fern\'andez D., Rosas-Ortiz O. and
Hussin V., {\em J. Phys. Conf. Ser.} {\bf 128} 012023 (2008)

\bibitem{Cru07}
Cruz~y~Cruz S., Negro J. and Nieto L.M., {\em Phys. Lett. A} {\bf
369} 400 (2007); Cruz~y~Cruz S., Negro J. and Nieto L.M., {\em J.
Phys.: Conf. Ser.} {\bf 128} 012053 (2008)

\bibitem{Cru08}
Cruz~y~Cruz S., Kuru S. and Negro J., {\em Phys. Lett. A} {\bf 372}
1391 (2008)

\bibitem{Souza}
de Souza Dutra A and de Oliveira J.A., \emph{Phys. Scr.} \textbf{78}
035009 (2008)




\end{thebibliography}
\end{document}